\documentclass[aps,12pt,superscriptaddress,preprint]{revtex4}

\usepackage[dvips]{graphicx}
\usepackage{amsmath,amssymb,amsthm,bm}

\newcommand{\R}{\mathbb{R}}
\newcommand{\Z}{\mathbb{Z}}

\begin{document}

\preprint{IUHET-518}

\vspace*{0.75in}

\title{Higher order Josephson effects}

\author{Roman V. Buniy}
\email{roman.buniy@gmail.com}
\affiliation{Physics Department, Indiana University, Bloomington, IN 47405}

\author{Thomas W. Kephart}
\email{tom.kephart@gmail.com}
\affiliation{Department of Physics and Astronomy, Vanderbilt
University, Nashville, TN 37235} 

\date{August 13, 2008}

\begin{abstract}
  Gaussian linking of superconducting loops containing Josephson
  junctions with enclosed magnetic fields give rise to interference
  shifts in the phase that modulates the current carried through the
  loop, proportional to the magnitude of the enclosed flux. We
  generalize these results to higher order linking of a
  superconducting loop with several magnetic solenoids, and show there
  may be interference shifts proportional to the product of two or
  more fluxes.
\end{abstract}


\maketitle

\section{Introduction}

Interference effects, both constructive and destructive, are mainstays
for distinguishing between quantum and classical phenomena. Examples
include interfering scattering amplitudes, both bosonic and fermionic,
formation of condensates, entanglement, etc. The Aharonov-Bohm
effect~\cite{Aharonov:1959fk}, describes the self-interference of a
charged particle that can travel along two semiclassical paths whose
combined path is gaussian linked with a magnetic solenoid carrying
flux $\Phi$. The measurable phase shift is $\phi\propto\Phi$. We have
argued in Ref.~\cite{Buniy:2006tq} that there could exist
generalizations to cases of higher order linkings. The simplest
example is a Borromean ring arrangement where the semiclassical path
corresponds to one ring, which has higher order linking with two flux
tubes carrying fluxes $\Phi_1$ and $\Phi_2$, which make up the other
two rings. We found the phase shift in this system is $\phi \propto
\Phi_1 \Phi_2$. Higher order cases were explored in
Ref.~\cite{Buniy:2006tr,Buniy:2006ts} and shown to be related to
commutator algebras of homotopy generators of the configuration space
$\R^3\backslash\{T_1\cup T_2\}$, where $T_1$ and $T_2$ are the tubes
containing the fluxes. The same general logic can be applied to
systems of superconductors, Josephson junctions, and magnetic fluxes
where the Josephson effect can arise \cite{josephson}. Here we will
study interference in a symmetric arrangement of two identical
semicircular superconductors joined by two identical Josephson
junctions and derive the response of such systems. We conclude with a
discussion of possible applications.

In the case of gaussian linking of a loop of superconductor with a
magnetic solenoid, the Mercereau effect~\cite{Mercereau} is due to the
phase change in the macroscopic wave function, which is in turn
related to the currents in the superconducting components. The effect
is due to the presence of a vector potential $\bm{A}$, which is the
fundamental object responsible for the phase change. Exploration of
higher order linking is again due to the presence of a vector
potential but in these instances it requires careful choices of gauge.

\section{The Josephson effect}

It will be sufficient for our purposes to consider a macroscopic model
of superconductors. Following Feynman~\cite{FeynmanLectures}, we
approximate the superconductors coupled via a Josephson junction as a
two-level system. Let $\psi_1$ and $\psi_2$ be the states, $E_1$ and
$E_2$ the energy levels of the superconductors. The Schrodinger
equation for the coupled system of the two superconductors becomes
\begin{align}
  i\hbar(\partial\psi_1/\partial t)
  &=E_1\psi_1+K\exp{(i\phi)}\psi_2,\\ i\hbar(\partial\psi_2/\partial
  t) &=E_2\psi_2+K\exp{(-i\phi)}\psi_1,
\end{align}
where $K$ is the coupling energy and $\phi$ is a phase, which arises
from the most general hermitian hamiltonian $2\times 2$ matrix. The
dependence of $\phi$ on the vector potential $\bm{A}$ can be found
from gauge invariance considerations. For a gauge transformation
$\bm{A}\mapsto \bm{A}+\bm{\nabla} f$,
$\psi_1\mapsto\psi_1\exp{(iqf/2\hbar c)}$,
$\psi_2\mapsto\psi_2\exp{(-iqf/2\hbar c)}$ with an arbitrary function
$f$ of space coordinates, we find $\phi\mapsto\phi+qf/\hbar c$, from
which it follows that $\phi=(q/\hbar c)\int \bm{A}\cdot d\bm{x}$. Here
$q=2e$ is the charge of an electron pair.

After the substitutions $\psi_1=\vert\psi_1\vert\exp{(i\theta_1)}$ and
$\psi_2=\vert\psi_2\vert\exp{(i\theta_2)}$, the Schrodinger equation
becomes
\begin{align}
  \hbar(\partial\vert\psi_1\vert^2/\partial t)
  &=2K\vert\psi_1\vert\vert\psi_2\vert\sin{\theta},\\
  \hbar(\partial\vert\psi_2\vert^2/\partial t)
  &=-2K\vert\psi_1\vert\vert\psi_2\vert\sin{\theta},\\
  -\hbar\vert\psi_1\vert(\partial\theta_1/\partial t)
  &=E_1\vert\psi_1\vert+K\vert\psi_2\vert\cos{\theta},\\
  -\hbar\vert\psi_2\vert(\partial\theta_2/\partial t)
  &=E_2\vert\psi_2\vert+K\vert\psi_1\vert\cos{\theta},
\end{align}
where $\theta=\phi+\theta_2-\theta_1$. The current from superconductor
$1$ to superconductor $2$, which is equal to minus the current from
superconductor $2$ to superconductor $1$, is thus
$I=(2K/\hbar)\vert\psi_1\vert\vert\psi_2\vert\sin{\theta}$. (In a
self-consistent computation, a current from a battery which connects
the two superconductors is also included. The result for the
superconducting current is precisely $I$; see, for example,
Ref.~\cite{Ohta}.) The electron densities in the two superconductors
are approximately equal and independent of time; let $\rho$ be this
common constant. This gives $I=I_0\sin{\theta}$, where
$I_0=2K\rho/\hbar$. Integrating the phase equations, we find
\begin{align}
  \theta(t)=\phi+\theta(0)+\hbar^{-1}\int_0^t dt'\,(E_1(t')-E_2(t')).
\end{align}
The quantity $(E_1(t)-E_2(t))/q$ represents an electric potential
applied to the junction. The dc and ac Josephson
effects~\cite{josephson} arise for $\vert E_1(t)-E_2(t)\vert\ll K$ and
$\vert E_1(t)-E_2(t)\vert\gg K$, respectively.

\begin{figure}[ht]
  \includegraphics[width=6cm]{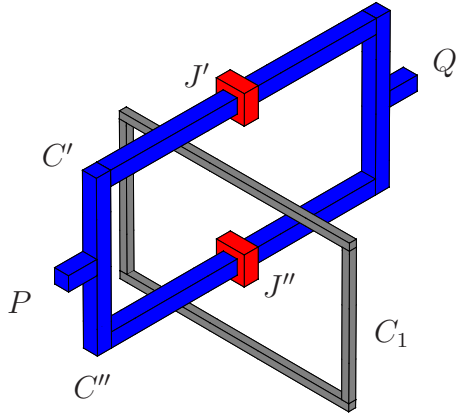}
  \caption{\label{figure-josephson-1} A diagram of an experimental
  setup for the detection of the Josephson effect. $C'$ and $C''$ are
  paths from the point $P$ to the point $Q$ through the
  superconductors with the Josephson junctions $J'$ and $J''$ and the
  total current $I$ from $P$ to $Q$. $C_1$ is the magnetic solenoid
  carrying flux $\Phi_1$. The Josephson effect (for a review see
  Ref.~\cite{RMP}) is due to the first order (gaussian) linking of the
  closed curves $C=C'C^{\prime\prime -1}$ and $C_1$.}
\end{figure}

Our interest here is in the Josephson effect with zero potential
across the junction, $E_1(t)-E_2(t)=0$, and nonzero magnetic field
constrained to the opening of the superconducting ring with two
Josephson junctions; see Fig.~\ref{figure-josephson-1}. Let $\theta'$
and $\theta''$ be the phase changes due to the vector potential
$\bm{A}_1$ of the currents through the junctions $J'$ and $J''$. The
phase changes from the point $P$ to the point $Q$ along the paths $C'$
and $C''$ are
\begin{align}
  \phi'&=\theta'+(q/\hbar c)\int_{C'} \bm{A}_1 \cdot d\bm{x},\\
  \phi''&=\theta''+(q/\hbar c)\int_{C''} \bm{A}_1 \cdot d\bm{x}.
\end{align}
Since the wave function is single valued, this requires
$\phi'=\phi''$, and so we find
$\theta''-\theta'=2\pi\Phi_1/\Phi_0$. Here $\Phi_1=\oint_C
\bm{A}_1\cdot d\bm{x}$ is the flux due to the solenoid along $C_1$
passing through a surface spanned by a closed curve
$C=C'C^{\prime\prime -1}$ and $\Phi_0=2\pi\hbar c/q$ is the flux
quantum. The total current from the point $P$ to the point $Q$ is
\begin{align}
  I=I_0\sin{(\tfrac{1}{2}(\theta'+\theta''))}\cos{(\pi\Phi_1/\Phi_0)}.
\end{align}
For a fixed value of $\Phi_1$, the corresponding maximal total current
is
\begin{align}
  I_{\textrm{max}}
  =I_0\vert\cos{(\pi\Phi_1/\Phi_0)}\vert\label{Imax-order-1},
\end{align}
which itself has maxima when $\Phi_1=n\Phi_0$, $n\in\Z$.

The flux is actually $\Phi_1=\Phi_{1,\textrm{ext}}+LI$ where
$\Phi_{1,\textrm{ext}}$ is the external flux through the loop, $L$ is
the self-inductance, but here and in what follows we assume $L$ is
negligible. (We have made a number of simplifying assumptions, for
example, that self-inductance of SQUID components are negligible, none
of which, if relaxed, affect our basic conclusions.)

We will call the phenomena reviewed in this section first order
Josephson effects to distinguish them from their generalizations which
we now proceed to describe.

\section{The second and higher order Josephson effects}

\begin{figure}[ht]
  \includegraphics[width=7cm]{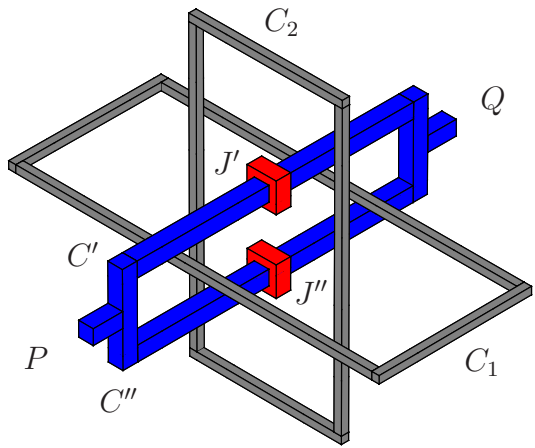}
  \caption{\label{figure-josephson-2} A diagram of an experimental
  setup for the detection of the second order Josephson effect. $C_1$
  and $C_2$ are the magnetic solenoids. $C'$ and $C''$ are paths from
  the point $P$ to the point $Q$ through the superconductors connected
  by Josephson junctions $J'$ and $J''$. The total current from $P$ to
  $Q$ is $I$. $C_1$ and $C_2$ are magnetic solenoids carrying fluxes
  $\Phi_1$ and $\Phi_2$. The second order Josephson effect is due to
  the second order linking of the set of three closed curves
  $C=C'C^{\prime\prime -1}$, $C_1$ and $C_2$.}
\end{figure}

Now consider the case where we have two solenoids carrying magnetic
fluxes $\Phi_1$ and $\Phi_2$ and whose center lines run along $C_1$
and $C_2$, and a superconducting ring along the closed curve
$C=C'C^{\prime\prime -1}$ with two Josephson junctions $J'$ and $J''$
in parallel as shown in Fig.~\ref{figure-josephson-2}. The two
solenoids and the superconducting ring are in a Borromean
rings~\cite{Rolfson} configuration. Note that in this arrangement
neither $C_1$ nor $C_2$ has gaussian linking with the superconducting
ring $C$, nor do $C_1$ and $C_2$ link with each other. However, the
set of three rings $C,C_1,C_2$ is indeed linked. This second order
linking and its higher order generalizations is what will lead to our
results. In other words, we will find that even though our system
lacks first order (gaussian) linking, a phase difference can still
exist upon traveling around the superconductor. To find this phase, we
must choose a gauge that detects it. Such a gauge is
\cite{Buniy:2006tq} $\bm{A}_{12}
=\tfrac{1}{2}k_2(\gamma_1\bm{A}_2-\gamma_2\bm{A}_1)$. Here $\bm{A}_1$
and $\bm{A}_2$ are the vector potentials due to the solenoids along
$C_1$ and $C_2$, the quantities $\gamma_1$ and $\gamma_2$ are defined
by
\begin{align}
  \gamma_j=\delta_j+(q/\hbar c)\int_\Gamma\bm{A}_j\cdot d\bm{x},
\end{align}
where $\delta_1$ and $\delta_2$ are constants, and $\Gamma$ is a path
that runs along $C$. The quantity $k_2$ is a normalization constant,
the value of which we discuss below.

Using $\bm{A}_{12}$ in the phase integral and following computations
in Ref.~\cite{Buniy:2006tq}, we find $\theta''-\theta'=4\pi^2
k_2\Phi_1\Phi_2/\Phi_0^2$. The total current from point $P$ to point
$Q$ is
\begin{align}
  I =I_0\sin{(\tfrac{1}{2}(\theta'+\theta''))}\cos{(2\pi^2
  k_2\Phi_1\Phi_2/\Phi_0^2)}.
\end{align}For fixed values of
$\Phi_1$ and $\Phi_2$, the maximal total current flowing in the
superconductor is
\begin{align}
  I_{\textrm{max}} =I_0\vert\cos{(2\pi^2
  k_2\Phi_1\Phi_2/\Phi_0^2)}\vert. \label{Imax-order-2}
\end{align}

The smallest value of the constant $k_2>0$ for which the fluxes
$\Phi_1=m_1\Phi_0$, $\Phi_2=m_2\Phi_0$, where $m_1,m_2\in\Z$, lead to
maxima of the quantity $I_\textrm{max}$ is $k_2=(2\pi)^{-1}$. This is
precisely the value we obtained in Ref.~\cite{Buniy:2006tq} by
imposing an analog of the Dirac string condition on the second order
phase for the Aharonov-Bohm effect. Nevertheless, the value of $k_2$
must ultimately be determined by experiment.

Also, for the value $k_2=(2\pi)^{-1}$, if either $\Phi_1$ or $\Phi_2$
is equal to $\Phi_0$ or $-\Phi_0$, then in terms of the other flux,
appropriately relabeled, the expression~\eqref{Imax-order-2} for the
second order $I_{\textrm{max}}$ reduces to the
expression~\eqref{Imax-order-1} for the first order $I_\textrm{max}$.

An essential feature of the above result is that the current $I$ is a
periodic function of the quantity $\pi k_2\Phi_1\Phi_2/\Phi_0^2$ with
period $1$. We can also derive this property by modifying the method
used by Block~\cite{Block} for the first order Josephson effect as
follows.

The total gauge potential includes internal and external parts,
$\bm{A}=\bm{A}_\textrm{in}+\bm{A}_\textrm{ext}$, the external magnetic
field being due to the external sources. Assuming that the external
field $\bm{\nabla}\times\bm{A}_\textrm{ext}$ vanishes inside the
superconductors, we can write
$\bm{A}_\textrm{ext}=\bm{\nabla}\gamma_\textrm{ext}$. As a result,
$\bm{A}$ is a gauge transformation of $\bm{A}_\textrm{in}$, and so
\begin{align}
  \psi(\bm{A}_\textrm{in}+\bm{A}_\textrm{ext})
  =\psi(\bm{A}_\textrm{in})\exp{(iq\gamma_\textrm{ext}/\hbar c)}.
\end{align}
Since $\psi(\bm{A})$ is single valued, we find that
$\psi(\bm{A}_\textrm{in})$ is multiplied by the factor $\exp{(-4i\pi^2
k_2\Phi_1\Phi_2/\Phi_0^2)}$ after the charge $q$ travels around a
closed curve $C$. This factor is a periodic function of $\pi
k_2\Phi_1\Phi_2/\Phi_0^2$ with period $2$. This implies the same
periodicity property for the wave function $\psi(\bm{A}_\textrm{in})$
and the energy $E$. Assuming time reversal symmetry as in
Ref.~\cite{Block}, we find that the free energy and thus the current,
which is given by minus the derivative of the free energy with respect
to the external flux, are both periodic functions of $\pi
k_2\Phi_1\Phi_2/\Phi_0^2$ with period $1$, in agreement with the
result proved earlier.

More generally, it is straightforward to arrange $n$ solenoids with
the fluxes $\Phi_1,\ldots,\Phi_n$ and a superconducting ring in such a
way that they are linked with nonzero $n{\textrm{th}}$ order
linking~\cite{Rolfson}. We similarly find the phase difference
\begin{align}
  \theta''-\theta'=(2\pi)^n k_n\Phi_1\cdots\Phi_n/\Phi_0^n,
\end{align}
the current
\begin{align}
  I =I_0\sin{(\tfrac{1}{2}(\theta'+\theta''))}
  \cos{(\tfrac{1}{2}(2\pi)^n k_n\Phi_1\cdots\Phi_n/\Phi_0^n)},
\end{align}
and its maximal value for fixed values of $\Phi_1,\ldots,\Phi_n$,
\begin{align}
  I_{\textrm{max}} =I_0\vert\cos{(\tfrac{1}{2}(2\pi)^n
  k_n\Phi_1\cdots\Phi_n/\Phi_0^n)}\vert
  \label{Imax-order-n}.
\end{align}
Other properties of these systems can be investigated. 

The smallest value of the constant $k_n>0$ for which the fluxes
$\Phi_j=m_j\Phi_0$, where $m_j\in\Z$, lead to maxima of the quantity
$I_\textrm{max}$ is $k_n=(2\pi)^{1-n}$. Again this is precisely the
value we obtained in Refs.~\cite{Buniy:2006tq,Buniy:2006tr} by
imposing an analog of the Dirac string condition on the phase for the
$n{\textrm{th}}$ order Aharonov-Bohm effect. Nevertheless, as pointed
out with the $k_2$ case above, the value of $k_n$ must ultimately be
determined by experiment.

Also, for the value $k_n=(2\pi)^{1-n}$, if one of the fluxes is equal
to $\Phi_0$ or $-\Phi_0$, then in terms of the remaining fluxes,
appropriately relabeled, the expression~\eqref{Imax-order-n} for the
$n{\textrm{th}}$ order $I_{\textrm{max}}$ reduces to the analogous
expression~\eqref{Imax-order-n} for the $(n-1){\textrm{st}}$ order
$I_\textrm{max}$.

Similar to the case $n=2$ above, we can modifying the method used by
Block and prove that for any $n$ the current $I$ is a periodic
function of the quantity $\tfrac{1}{2}(2\pi)^{n-1}
k_n\Phi_1\cdots\Phi_n/\Phi_0^n$ with period $1$.

\section{Conclusion}

We generalize the Josephson effect to its higher order analogs in
which a superconducting loop links with several magnetic solenoids and
the resulting interference shifts are proportional to the product of
two or more fluxes.

One can conceive of a number of applications for devices build to take
advantage of higher order linking.  Such a system could be less
invasive than first order devices because it could keep the SQUID some
distance from an experimental sample. Possible applications include
both rf and dc SQUIDs that measure higher order linking of multiple
fluxes. Under some circumstances such devices could be useful in
measurements of complex biological systems, or any systems where
direct gaussian linking of a magnetic flux with a SQUID is
impractical, but where higher order linking is possible. For example,
one could have a system of (i) a fixed but adjustable flux tube, i.e.,
a solenoid; (ii) an unknown flux to be measured, and (iii) a SQUID. If
the three components can be arranged to have higher-order linking,
then the unknown flux could be measured, even though it has no
gaussian linking with the SQUID.

\begin{acknowledgments}

The work of RVB was supported by DOE grant number DE-FG02-91ER40661
and that of TWK by DOE grant number DE-FG05-85ER40226.

\end{acknowledgments}

\end{document}